\documentclass[prl,reprint,superscriptaddress,showpacs,preprintnumbers,amsmath,amssymb, hidelinks]{revtex4-1}

\pdfoutput=1

\usepackage{amsmath}
\usepackage{graphicx}
\usepackage{natbib}
\usepackage{verbatim}
\usepackage[withbib, all]{authorindex}	
\usepackage[usenames]{color}
\usepackage{bm}
\usepackage{hyperref}

\hypersetup{
    colorlinks, linkcolor={blue},
    citecolor={blue}, urlcolor={blue},
}

\begin{document}

\title{Experimental Demonstration Of Scanned Spin-Precession Microscopy}

\author{V. P. Bhallamudi}\email{bhallamudi.1@osu.edu}
\affiliation{Department of Physics, The Ohio State University, Columbus, Ohio 43210, USA}

\author{C. S. Wolfe}
\affiliation{Department of Physics, The Ohio State University, Columbus, Ohio 43210, USA}

\author{V. P. Amin}
\affiliation{Department of Physics and Astronomy, Texas A\&M University, College Station, Texas, USA}

\author{D. E. Labanowski}
\affiliation{Department of Physics, The Ohio State University, Columbus, Ohio 43210, USA}

\author{ A. J. Berger}
\affiliation{Department of Physics, The Ohio State University, Columbus, Ohio 43210, USA}

\author{D. Stroud}
\affiliation{Department of Physics, The Ohio State University, Columbus, Ohio 43210, USA}

\author{J. Sinova}
\affiliation{Department of Physics and Astronomy, Texas A\&M University, College Station, Texas, USA}

\author{P. C. Hammel}\email{hammel@physics.osu.edu}
\affiliation{Department of Physics, The Ohio State University, Columbus, Ohio 43210, USA}

\date{\today}

\maketitle

\textbf{ 
Progress in spintronics has been aided by characterization tools tailored to certain archetypical materials \cite{insightIssue:Spintronics.Nat.Material1, Zutic:2004p477, Awschalom:2007, Jedema:2002p1359, Crooker:2005p1461, Lou:2007p121}. New device structures and materials \cite{  Sasaki:2010p1408,  Pi:2010p1502, appelbaum_electronic_2007, Koo:2009p1412, Dash:2009p1462, Jungwirth:Kamil.PRL.2012} will require characterization tools that are material independent, provide sufficient resolution to image locally-varying spin properties and enable subsurface imaging. Here we report the demonstration of a novel spin-microscopy tool based on the variation of a global spin-precession signal in response to the localized magnetic field of a scanned probe. We map the local spin density in optically pumped GaAs from this spatially-averaged signal with a resolution of 5.5\,$\mu$m. This methodology is also applicable to other spin properties and its resolution can be improved. It can extend spin microscopy to device structures not accessible by other techniques, such as buried interfaces and non-optically active materials, due to the universal nature of magnetic interactions between the spins and the probe.
}

Precession in a magnetic field is a hallmark characteristic of a magnetic moment, which makes it a powerful discriminant of spin-related phenomena in spintronic studies; this is exploited in Hanle effect measurements \cite{meier_optical_1984}. It can also result in loss of spin information in the presence of unwanted magnetic fields, especially spatially varying ones. However, in this report we show that the spin precession due to the spatially inhomogeneous magnetic field of a micromagnetic probe\,($\mu P$) can be employed to encode local spin information into a spatially integrated measurement. This information can then be decoded using standard deconvolution techniques.

To demonstrate our technique we first measure a signal proportional to the globally-averaged spin density, $\Sigma$, as a function of the $\mu P$'s position. From this measurement we can then obtain a quantity proportional to the unperturbed spin density\,(i.e. the steady state density in the absence of the probe), ${\bm \rho}$, by deconvolving the signal from a theoretical or experimentally-determined Precessional Response Function\,(PRF). The PRF captures the response of the spins to the magnetic environment they experience. We repeat this process in the presence of an applied uniform transverse magnetic field to further demonstrate our understanding of this microscopy tool.
\begin{figure*}
\center{\includegraphics[width=0.75\linewidth]{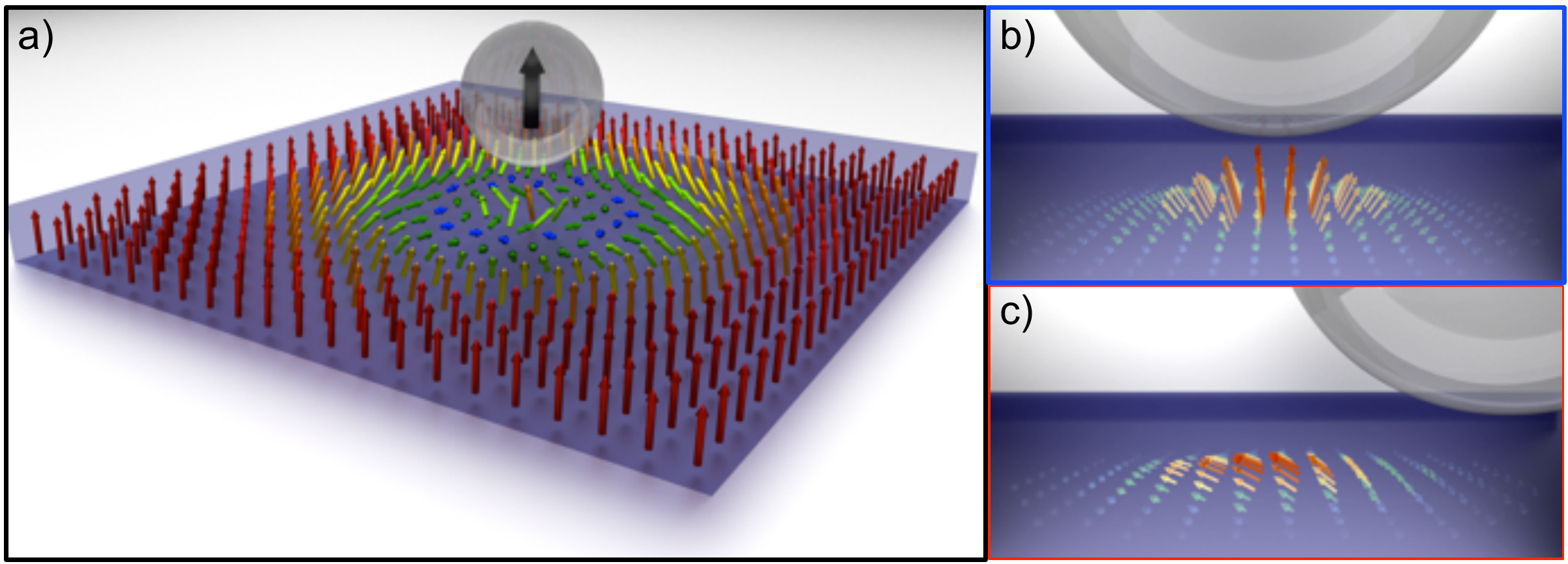}}
\caption{The simulations shown illustrate the key physics underlying Spin Precession Imaging; the inhomogeneous field of a micromagnetic probe $\mu P$ \,(sphere with black arrow, indicating the magnetization direction) generates a well-understood spatially-varying dephasing of spins. Spins that are injected into a semiconducting sample reach a steady-state density resulting from a combination of the local spin properties, the injected density, and the magnetic field due to the $\mu P$. The arrows in all figures represent the steady-state spin density vectors $\textbf{S}$.  The color scale represents the $\hat{\bf z}$ component\,(parallel to the orientation of the injected spin) of the spins. \textbf{a)} Demonstrates the case of a spatially-uniform injection density and  highlights the full spatial-extent of the $\mu P$'s influence. Strong transverse fields from the probe dephase spins in an annular disk centered beneath the probe. Parallel fields immediately below the probe protect spins from dephasing. \textbf{b)} and \textbf{c)} show simulations identical to panel \textbf{a)} except for the case of a spatially confined injection profile, corresponding to two different $\mu P$ positions. Panels \textbf{b} and \textbf{c} correspond to the blue and red dots in Fig.~\ref{fig:gaussianBeamData_1}\,\textbf{c} respectively. 
}
	\label{fig:fig1_1}
\end{figure*}
\begin{figure*}
\center{\includegraphics[width= 0.75\linewidth]{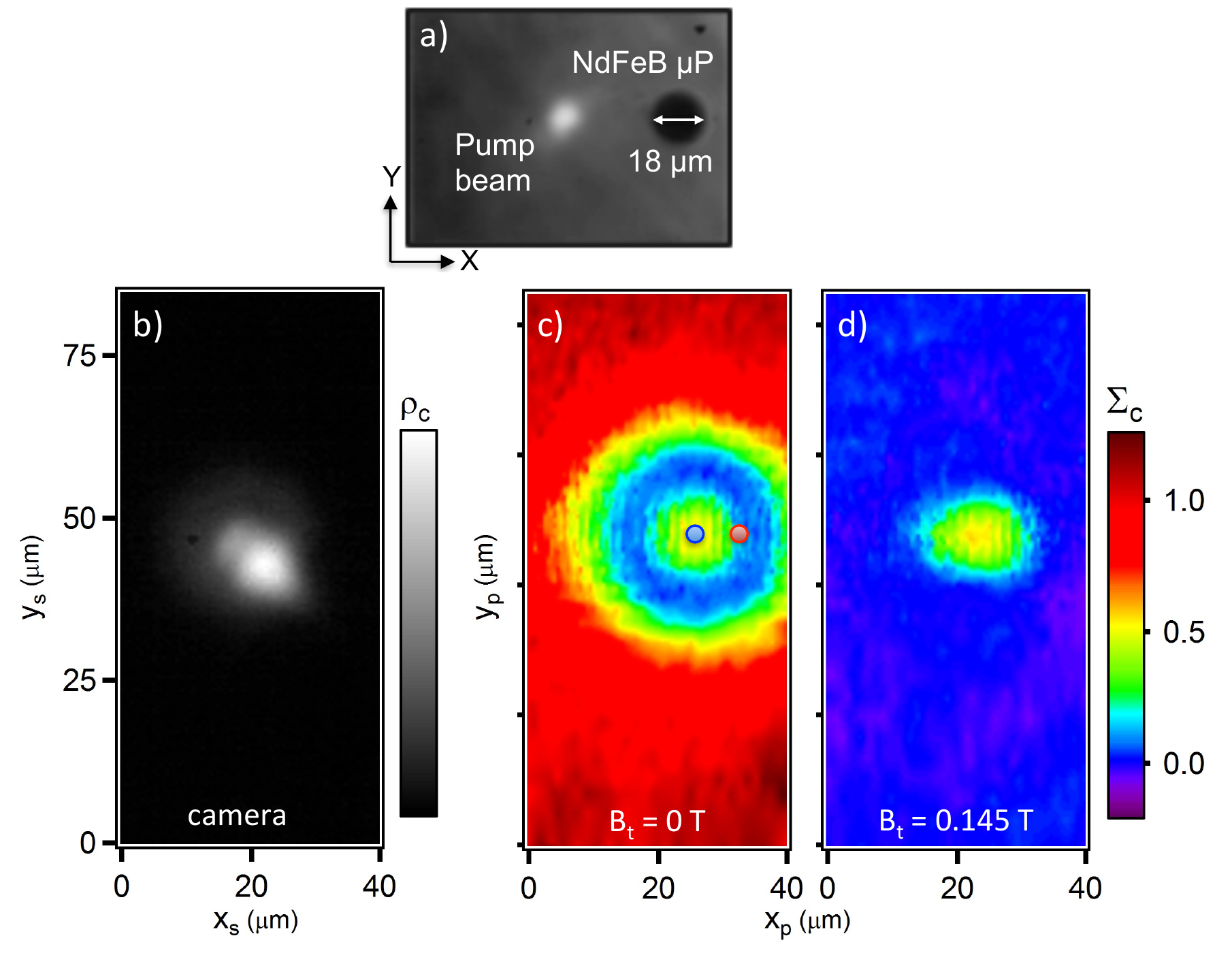}}
\caption{Spatial maps of the injection beam profile from the optical pump, the injected spin profile and the response of the global spin signal to probe position relative to the pump. \textbf{a)} Spin-insensitive camera image showing the PL spot created by the pump beam in a $1\,\mu$m thick GaAs membrane. Also visible is the the NdFeB micromagnetic probe $\mu P$, magnetized along the sample normal and glued to the membrane on the side opposite to that of the pump beam.  An additional uniform magnetic field $B_t\,\hat{\bf x}$ may be applied.  In the experiment the pump beam is scanned relative to the $\mu P$ using an objective mounted on translation stages. \textbf{b)} Camera image of PL that corresponds to\,(and is proportional to) injection profile $\rho_c$. \textbf{c)} The measured spin signal $\Sigma_c$ (colorbar) corresponding to the injection profile from panel \textbf{b}, at $B_t = 0$.  The location of each pixel corresponds to the relative position between the $\mu P$ and the pump beam; a blue and red dot are included to provide two example cases. The location of the blue dot corresponds to a $\mu P$ position directly above the injection beam; this configuration results in a spin density mostly pointing along the $\hat{\bf z}$ direction (Fig.~\ref{fig:fig1_1}\textbf{b}) and gives a large signal. The red dot corresponds to a $\mu P$ position 8 $\mu$m away from the center of the beam; spins precess away from the $\hat{\bf z}$\ direction due to large perpendicular fields (Fig.~\ref{fig:fig1_1}\textbf{c}) and thereby reduce the signal. \textbf{d)} $\Sigma_c$ for a large $B_t = 0.145$\,T. 
}
	\label{fig:gaussianBeamData_1}
\end{figure*}

We generate ${\bm \rho}= \rho({\bf r}_s)\hat{\bf z}$, in a GaAs membrane via optical pumping \cite{meier_optical_1984}, where ${\bf r}_s = (x_s, y_s, 0)$ is the position within the sample which we treat as two dimensional. Due to precession the steady state spin density in the presence of any magnetic field will be different from $\bm \rho$, and we will denote it by $\bf S$.  We measure a globally averaged spin-PhotoLuminescence\,(PL) signal \cite{meier_optical_1984}, $\Sigma \propto \int_{-\infty}^\infty  S_z\mathrm{d}^2{\bf r}_s$, where $S_z$ refers to the $\hat{\bf z}$ component of $\bf S$. 

This signal is measured while the pump laser is scanned relative to the $\mu P$ in the $\hat{\bf x}$ and $\hat{\bf y}$ directions. For our uniform sample, this is equivalent to scanning the $\mu P$ relative to a fixed pump, and henceforth we will regard this to be the case. The field ${\bf B}_p$ of the $\mu P$ modifies the precession behavior of the spins\,(see Fig.~1). This field depends on $\mu P$'s position. A spatially uniform transverse magnetic field, $B_t\hat{\bf x}$, may also be applied to further tailor this precession.  More details of the experimental set-up and measurement techniques are presented in the Supplementary Information (SI).

Fig.~\ref{fig:gaussianBeamData_1}{\bf b} shows the measured PL intensity\,(which is proportional to $\rho)$ for a particular $\rho = \rho_c$, which will be used later for obtaining the PRF. Panels {\bf c} and {\bf d} show the corresponding $\Sigma_c$, for $B_t = 0$\,T and $0.145$\,T respectively, as a function of the $\mu P$'s position, ${\bf r}_p = (x_p, y_p, z_p)$.  

An expression for $\Sigma$ in the limit of small diffusion\,(which our data shows is a reasonable approximation for this experiment; more details of the derivation in the SI) is given by: 
\begin{equation}
\begin{aligned}[b]
\Sigma({\bf r}_p, B_t) 	&\propto \int_\infty^\infty  S_z({\bf r}_s, {\bf r}_p)\mathrm{d}^2{\bf r}_s\\
					&\propto \int_\infty^\infty H_B({\bf R}, B_t) \rho({\bf r}_s) \mathrm{d}^2{\bf r}_s\\
					&= H_B(\textbf{R}, B_t) \ast \rho(\textbf{r}_s)
	\label{eq:thetab}
\end{aligned}
\end{equation}

where, $ \ast$ represents a convolution, $\textbf{R} = \textbf{r}_p - \textbf{r}_s$ and
\begin{align*}
H_B &= \frac{1}{1 + \theta_B^2}\hspace{10pt}\mbox{and} &\theta_B^2(\textbf{R}, B_t) &= \frac{(\gamma\tau_s B_\perp)^2}{1 + (\gamma \tau_s B_\|)^2}.
\label{eq:thetab1}	
\end{align*}

$H_B$ is the PRF and is the signal which would be collected for a spatial delta function injection. $B_\perp$ and $B_\|$ are respectively the magnitudes of the perpendicular and parallel components of the total field, ${\bf B} = {\bf B}_p({\bf R})  + B_t\hat{{\bf x}}$, experienced by the spins. The components are defined with respect to $\hat{\bf z}$, the orientation of injected spins. The gyromagnetic ratio is denoted by $\gamma$ and  $\tau_s$ is the spin relaxation time. 

 As seen from the previous equations, $S_z$ is decreased by $B_\perp$ because it causes the spins to precess away from the injected direction, resulting in a dephasing of the ensemble.  On the other hand $B_\|$ keeps the spins from tipping away from the injection direction, resulting in a small $\theta_B$ and large $S_z$. We can view $\theta_B$ as an effective dephasing factor \cite{bhallamudi:JAP2012, jansen:InjModDopedSilicon.PhysRevB.2010}.

\begin{figure*} 
	 \center{\includegraphics[width=0.95\linewidth]{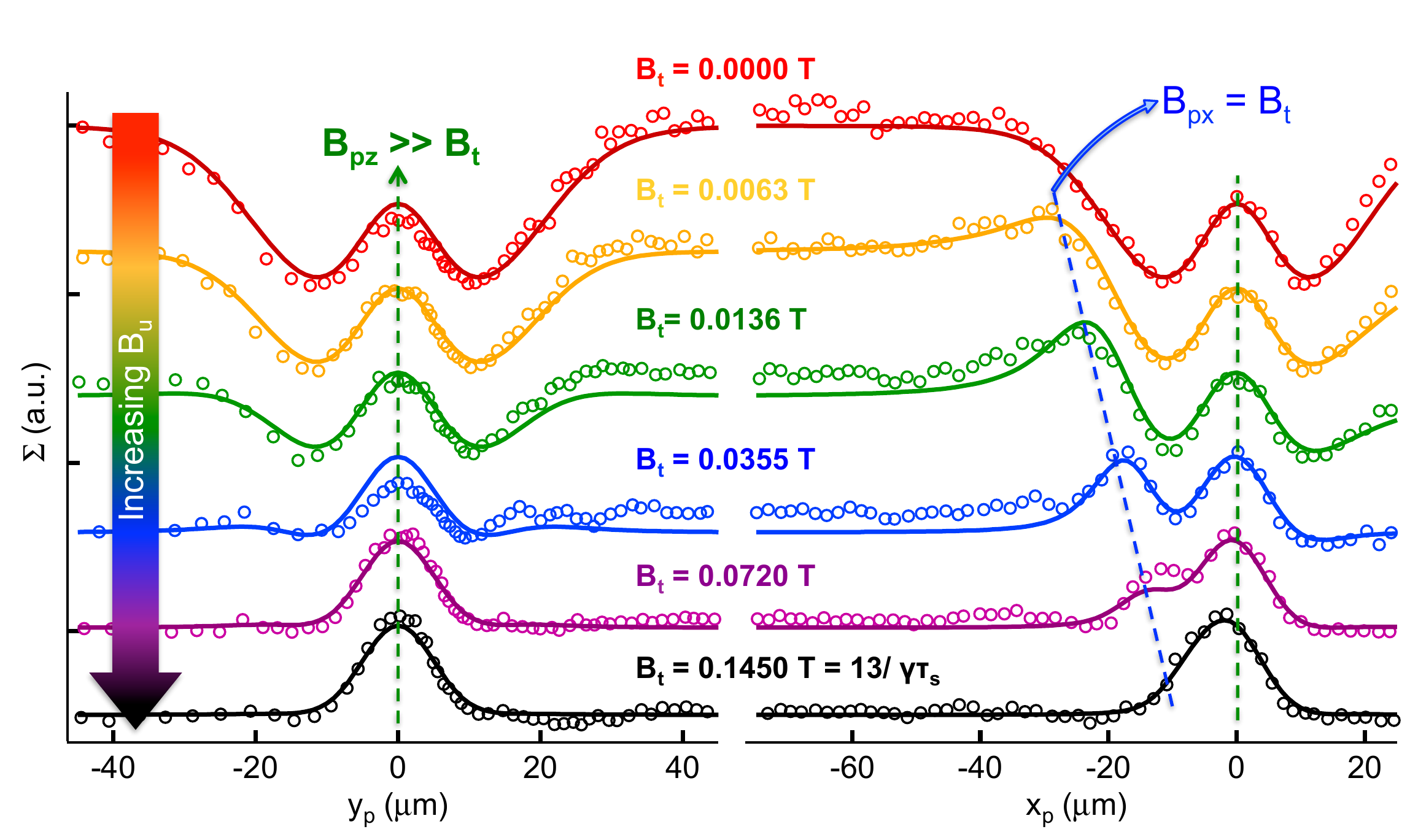}}
	\caption{Spin signal\,($\Sigma_c$, corresponding to $\rho_c$) for various $B_t$ plotted as a function of the relative position between the pump and the $\mu P$ along the $\hat{\bf y}$\ (left) and the $\hat{\bf x}$\,(right) directions.  The data at various $B_t$ are offset vertically for clarity of presentation. The open circles are experimental data while the solid lines represent fits obtained from Eqn.~1 using $m_z = 2\times 10^{-9}$\, J/T, $z_p$ = $8\,\mu$m, and $\tau_s = 2.33$\,ns. The spin lifetime $\tau_s$ is obtained from Hanle measurements (see SI) and the $\mu P$ radius ($z_p$) is taken to be $9\,\mu$m.  The peaks marked by the green dashed lines result when the peak in $\rho_c$ lies directly under the $\mu P$. The net parallel field directly under the $\mu P$ is mostly due to $B_{pz}$ (the $\hat{z}$ component of the $\mu P$'s field) and exceeds the net transverse field (mostly given by $B_t$); this results in a large $\Sigma$\,(Eqn.~1). The second set of peaks seen in scans along the $\hat{\bf x}$ direction\,(blue dashed line) occur when $B_{px} = -B_t$ ($B_{px}$ is the $\hat{\bf x}$ component of $\mu P$'s field).} 
	\label{fig:lineScansGaussianBeam}
\end{figure*}

 $B_\|$ and $B_\perp$ have distinct spatial variation in our experiments, and the consequences of their competing effects are evident in Fig.~\ref{fig:lineScansGaussianBeam}, where we show line scans (along $\hat{\bf x}$ and $\hat{\bf y}$) for several values of $B_t$. Also shown are fits obtained from Eqn.~1 in which the probe is modeled as a point dipole with a moment ${\bf m} = m_p \hat{\bf z}$ located a height $z_p$ above the sample (see Fig. 3 caption and the SI for more details).
 
 The peak, marked by vertical green dashed lines at $x$ or $y=0$, occurs when the $\mu P$ is located directly above the point of maximum injected spin density. At this point there is a maximum in $B_\parallel$\,($\sim$0.8 T) from the $\mu P$ that preserves $S_z$. When the pump is far from the $\mu P$, the signal decreases with increasing $B_t$ with a Lorentzian line shape\,(as expected in a Hanle measurement) whose half-width, $B_{1/2}$ = $(\gamma\tau_s)^{-1}$, is ~0.0111\,T for our experiments. 

 A second peak\,(blue dashed line) seen in the line scans along $\hat{\bf x}$ occurs where the $\hat{\bf x}$ component of the field from the $\mu P$ cancels $B_t$. As $B_t$ is increased, this point occurs closer to the $\mu P$ where its field is stronger.

 The fits indicate the effectiveness of Eqn.~1 in describing our data. The sensitivity of the global signal to spins at different locations relative to the $\mu P$, described by the convolution, forms the basis for imaging.

\begin{figure*}
\center{\includegraphics[width=0.95\linewidth]{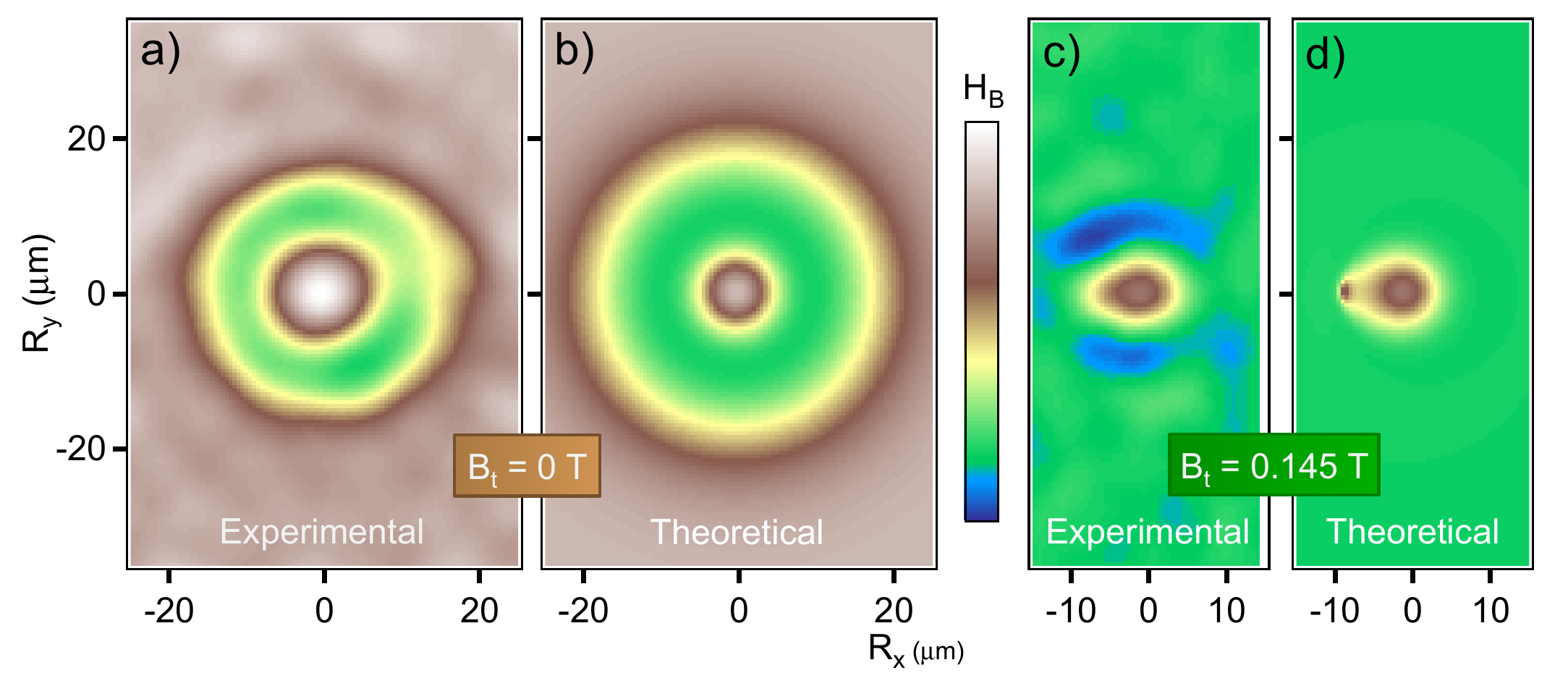}}
\caption{\textbf{a)} Precessional Response Function obtained through Wiener deconvolution of $\rho_c$\,(Fig.~\ref{fig:gaussianBeamData_1}\textbf{b}) from $\Sigma_c (B_t = 0)$\,(Fig.~\ref{fig:gaussianBeamData_1}\textbf{c}). \textbf{b)} Theoretically derived PRF obtained from Eqn.~1, using the dipole moment from the fits in Fig.~\ref{fig:lineScansGaussianBeam}. \textbf{c)} and \textbf{d)} similar experimental and theoretical PRFs for $B_t = 0.145 T$. The experimental PRF data has been normalized and offset to highlight the match between theory and experiment. 
}
	\label{fig:PRFs}
 \end{figure*}

 To obtain an unknown spin density, $\rho_u$, from our measured signal, the PRF needs to be known. The PRF can be obtained theoretically or from experimental data, if we have a known $\rho$. For the latter, we use the camera data shown in Fig~\ref{fig:gaussianBeamData_1}\textbf{d} as being proportional to $\rho_c$. Then $H_B(\textbf{R}, B_t) = \Sigma_c(\textbf{r}_p, B_t)\circledast \rho_c(\textbf{r}_s)$, where we use $\circledast$ to indicate a deconvolution process. We use the Wiener algorithm \cite{book:russ.ImageProcessing} to implement the deconvolution. The resulting experimental PRFs for both low and high $B_t$ are shown in Fig~\ref{fig:PRFs}. For comparison the theoretical PRFs are also shown; more details for obtaining them are presented in the SI.

\begin{figure*}
\center{\includegraphics[width=0.7\linewidth]{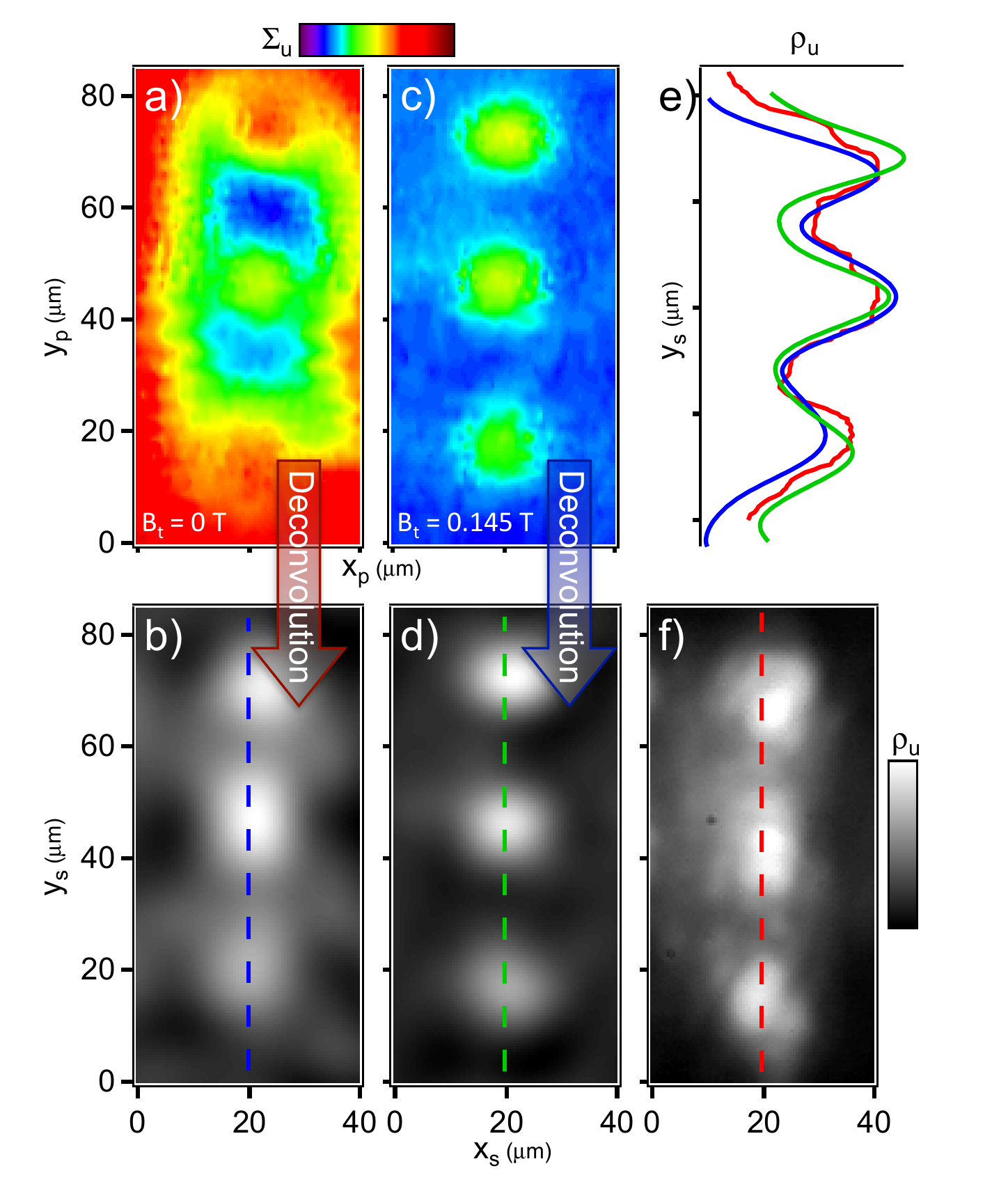}}
\caption{Obtaining the spatial variation of an unknown spin density from the measured signal $\Sigma$: \textbf{a)} Spin signal $\Sigma_u$ measured for a spin density profile, $\rho_u$\,(more information on how it was produced given in the SI), with $B_t = 0$\,T. \textbf{b)} An image of $\rho_u$ extracted from the Wiener deconvolution of $\Sigma_u$ (panel \textbf{a}) using the experimental $H_B(B_t = 0)$\,(Fig.~\ref{fig:PRFs}\textbf{a}) as the deconvolution kernel. \textbf{c)} and \textbf{d)} Similarly measured $\Sigma_u$ and extracted $\rho_u$ respectively for $B_t = 0.145$\,T. \textbf{e)} Line cuts of $\rho_u$ taken along the dashed lines in panels \textbf{b}\,(blue) and \textbf{d}\,(green). Also shown in red is a line cut from an independent camera image of $\rho_u$ shown in panel \textbf{f}.} 
	\label{fig:interferenceBeamData}
\end{figure*}

To test the fidelity of our imaging process, we now use the experimental PRF for an unknown $\rho_u$. The measured $\Sigma_u$, at low and high fields, is presented in panels \textbf{a} and \textbf{c} of Fig.~\ref{fig:interferenceBeamData}. We then extract $\rho_u(\textbf{r}_s) = \Sigma_u(\textbf{r}_p, B_t)\circledast H_B(\textbf{R}, B_t)$. The extracted spin densities are shown in panels \textbf{b} and \textbf{d} of the same figure. Also shown\,(panel \textbf{e}), for independent verification of our imaging technique, is a camera image for the PL ($\propto\rho_u$). The linecuts present a more quantitative comparison of the extracted and measured data.

 The ability to extract the spin density with both high and low $B_t$ shows the exclusion of spurious effects, such as reflectivity changes as a function of the $\mu P$'s position, in our data. Also, $B_t$ provides a knob to optimize the PRF to suit particular imaging needs. High field imaging might provide a more intuitive PRF for the case of global detection. Low-field imaging maybe more useful for non-local electrical devices, where a large transverse field would dephase spins before they reach the detector. 

The line scans shown in Fig.~\ref{fig:lineScansGaussianBeam} provide us a measure of the spatial resolution, $\zeta$, in our experiment. A Gaussian fitting of the narrowest lobe  gives us $\zeta \leq 5.5\,\mu$m. It is an upper bound that is being set by the feature size we are imaging. As in  magnetic resonance imaging, the magnetic field gradient, $\kappa$, sets the ultimate resolution in the absence of diffusion, $\zeta = B_{1/2}/\kappa$ \cite{bhallamudi:JAP2012}. Gradients of up to $\sim 4\times10^6$\,T/m have been reported recently \cite{r:tobaccomosaicvirus.pnas.2009}, which are at least an order of magnitude larger than in this experiment, and should enable much finer resolution. 

Resolution will be limited by diffusion and the unavoidable reduction of the spin signal as the detected volume shrinks. While diffusion can degrade the resolution, the spatial precessional response can be numerically analyzed to obtain valid and useful spatially resolved data. For large enough gradients, sub-diffusion length and sub-diffraction limit resolution should be achievable. Images are obtained in the presence of spin diffusion by MRI \cite{Tournier:DiffusionMRI.NeuroImage2004, Basser:DiffusionMRI.BiophysicslJr.1994, Kalvis:PASMRI.InverseProblems.2003}; this should be feasible for spin precession imaging as well. 

 In summary, we have demonstrated a new technique for imaging spin properties using the precessional response of spins to a micromagnetic probe's field. While we have imaged the variations in the spin density, the technique is more general since the response of the spins is sensitive to a variety of spin characteristics including spin lifetime and gyromagnetic ratio. Work is underway to generalize the technique presented here using scannable probes mounted on cantilevers. Due to the magnetic nature of interaction between the probe and the spins, which can extend through layers of a heterostructure and a few microns deep, this tool should enable subsurface imaging. This technique should be applicable to a wide variety of materials because it relies on proven spin polarization detection techniques. With optical detection it can enhance imaging resolution, and with electrical detection it can enable imaging where none exists at present.
 
 Funding for this research was provided by the Center for Emergent Materials at the Ohio State University, an NSF MRSEC (Award Number DMR-0820414). We wish to thank Cristian Cernov for creating the rendered images presented in this article and the SI.


\end{document}